# Ultrananocrystalline Diamond Film: Tribological Studies Against Metal and Ceramic Balls

Revati Rani[1], Niranjan Kumar[1*] and I-Nan Lin[2]

[1]Materials Science Group, Indira Gandhi Centre for Atomic Research, HBNI, Kalpakkam 603102 (INDIA)

[2]Department of Physics, Tamkamg University, Tamsui 25137 (TAIWAN)

## Abstract

In this present work, friction and wear behavior of UNCD thin film has been studied against three different sliding counterbodies i.e. $Al_2O_3$, SiC and steel balls. UNCD/steel sliding pair showed high friction coefficient in the beginning of the sliding process which decreased to the lower value after longer sliding passes. This behavior might be explained by oxidation mechanism. However, high friction value was observed in the case of UNCD/SiC and this was attributed to strong adhesive force acting across the sliding interfaces. Stable and low friction value was measured in UNCD/$Al_2O_3$ sliding combination. This was governed by chemically inert interfaces and graphite tribolayer formation.

## Introduction
Carbon based thin films are important for tribological applications due to their tunable microstructure and chemical properties. Nanocrystalline diamond (NCD) and ultra-nanocrystalline diamond (UNCD) thin films showed improved friction and wear resistance properties [1-3]. Understanding the tribological behavior of UNCD thin films against ceramics and metal counterbodies are technically important for various industrial applications. In this present study, tribological properties of UNCD film has been evaluated against three different ceramic and metallic balls such as $Al_2O_3$, SiC and 100Cr6 steel. Furthermore, detail characterizations have been carried out for the sliding combination of UNCD/$Al_2O_3$. Here, it is important to mention that tribolayer is one of the most significant aspects for the determination of friction and wear properties [2].

## Experimental Section
### Film Deposition and Characterization
The UNCD film was grown on mirror polished silicon (100) substrate in microwave plasma enhanced chemical vapor deposition (MPECVD, 2.45 GHz 6'' IPLAS-CYRANNUS) system. The UNCD film was deposited using $CH_4$ (1%)/Ar plasma media with a microwave power delivery of 1200 W, a pressure of 120 Torr and a substrate temperature of 450°C. The detailed deposition process is reported elsewhere [1].

The surface topography and roughness of film were analyzed by atomic force microscope (AFM, Park XE-100). The chemical properties of the film were examined by the micro-Raman spectrometer (Andor SR-500i-C-R, wavelength 532 nm). The hardness and elastic modulus of film were evaluated by nanoindentation (Triboindenter TI 950, USA) coupled with Berkovich diamond indenter with a tip having a radius of curvature 150 nm. A maximum load of 6 mN and a loading-unloading rate of 1.5 mN/minute were used. Rutherford Back Scattering technique with α particle (energy 3.8 MeV) as the projectile was used to have elemental analysis in UNCD film. The scattering angle was kept as 165°.

Tribological testing on UNCD thin film has been carried out using ball-on-disc micro-tribometer (CSM-Switzerland) working in linear reciprocating mode. The tribological properties of UNCD film have been studied against three different standard balls (diameter 6 mm). The standardized hardness value of $Al_2O_3$, SiC and 100Cr6 steel balls are 18 GPa, 26 GPa and 5 GPa, respectively. The normal constant load of 2 N, sliding speed of 4 cm/sec was used in each experiment. All tests were being run for the total sliding distance of 500 m which equals to 62,500 numbers of sliding cycles. The data acquisition rate of 10 Hz and stroke length of 4 mm were used for all experiments. The tests were carried out in ambient (dry and unlubricated conditions). During the tribology test, in-situ electrical contact resistance (ECR) measurement was performed to describe the tribolayer formation in UNCD/$Al_2O_3$ sliding combination.

The wear tracks and ball scars developed after tribology testing were examined using an optical microscope. Three-dimensional wear profiles in tracks were also obtained using 3D optical profilometer (Taylor Hobson). Grazing incidence X-ray diffraction (GIXRD, Bruker D8, Germany) was used for crystal structure analysis of UNCD film. For this measurement, incident angle of 6°, a step size of 0.1° and 2θ angular range of 30° to 80° was used. The same GIXRD measurement was carried out inside wear track also.

## Results and Discussion
The topography of UNCD film showed homogeneous distribution of diamond particles (Fig. 1a). The roughness value of the film is ~9 nm. In Raman spectra, a broad peak is deconvoluted into five segments (Fig. 1b). $v_1$, $v_2$ and $v_3$ peaks are associated to transpolyacetylene (*TPA*) which indicate the presence of UNCD structure [4].

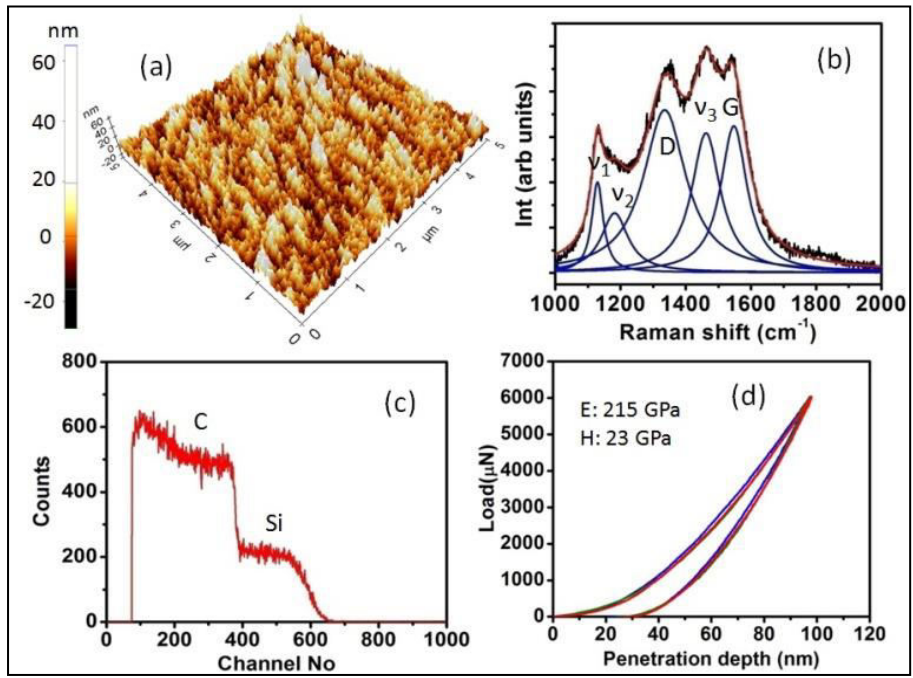

Fig. 1- (a) AFM image (b) Raman spectra (c) RBS spectra and (d) nano-indentation results of UNCD film

Other two peaks designated by G and D band correspond to graphite and an amorphous carbon phase. These phases are occupied by grain boundary of UNCD. RBS analysis indicates distinct signature of C and Si interfaces with film thickness 1.8 μm (Fig. 1c). Nano-indentation results showed elastic modulus and hardness values of 215 GPa and 23 GPa, respectively (Fig. 1d). The elastic recovery of the film was superior which showed less load-displacement hysteresis.

The friction results of UNCD film sliding against $Al_2O_3$, SiC and 100Cr6 steel balls are shown in Fig. 2a. Figure 2b shows in-situ electrical contact resistance (ECR) measurement in case of UNCD/$Al_2O_3$ sliding system. $Al_2O_3$ ball being chemically inert showed low friction value while sliding against UNCD film [5]. In this case, friction coefficient ~0.05 is maintained throughout the whole sliding process of 500 m after passing short run-in period of ~20 m. ECR measurement also follows the same trend as that of friction behavior. The trend of decrease in ECR is delayed as compared to friction coefficient. This could be associated with the sensitivity of the two different measurements i.e., ECR and friction coefficient. High friction at run-in regime is responsible to produce wear of the ball. This is shown in optical image (Fig. 3a).

SiC ball is also chemically stable but C-C strong chemical bond formation with UNCD film across sliding interfaces is possible during the sliding process. This may enhance the friction value (average 0.26) and in the response, wear of the ball is significant (Fig. 3b). In the case of UNCD/steel sliding combination, initially, high friction value was observed which gradually decreased to ~0.1 (Fig. 3c). Such frictional trend is associated with oxidation of sliding interfaces. In all three cases, wear of the film was negligible, indicating chemical and mechanical stability of the film. The sliding combination of UNCD/$Al_2O_3$ showed superior friction and wear resistance properties compared to other two. In this case, AFM and XRD analyses were performed inside the wear track. The results are shown in Fig. 4 and Fig. 5, respectively.

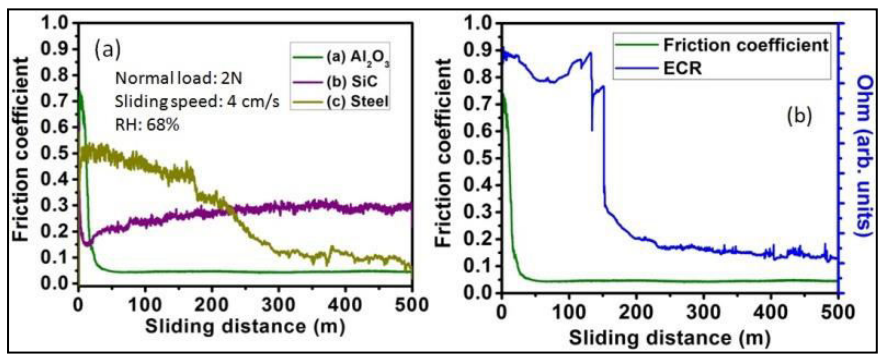

Fig. 2- (a) Friction results of UNCD film sliding against $Al_2O_3$, SiC and steel balls and (b) in-situ ECR measurement for UNCD/$Al_2O_3$ sliding system; Tribology parameters: Load: 2 N, Sliding speed: 4 cm/sec, Sliding distance: 500m

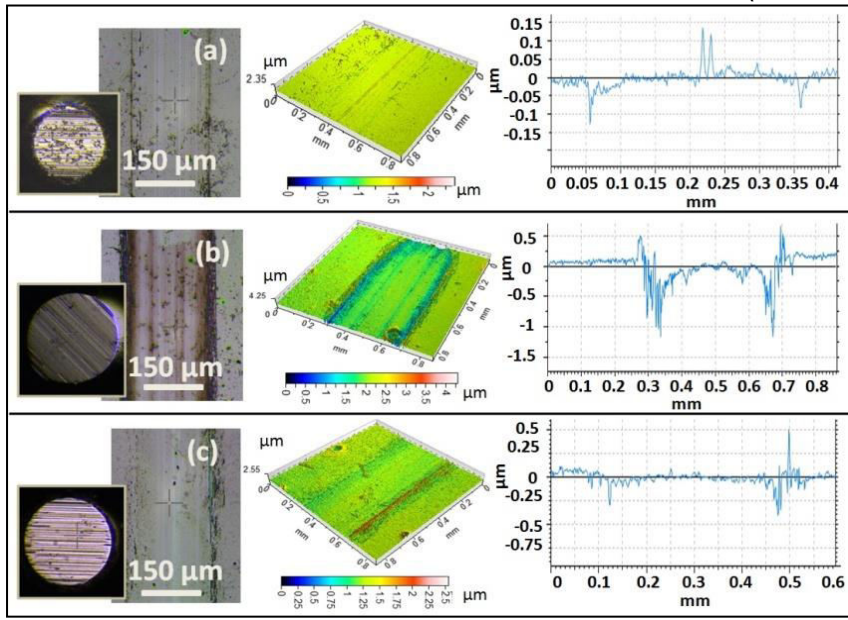

Fig. 3- Wear track and ball scar optical images along with 3D profile of wear track against (a) $Al_2O_3$, (b) SiC and (c) steel ball, respectively; Tribology parameter: Load: 2 N, Sliding speed: 4 cm/sec, Sliding distance: 500m

In AFM analysis, locally secondary phase formation is clearly seen which is quite similar to few layer graphite with the height of ~3 nm (Fig. 4d). In our previous study, similar results were obtained for the sliding combination of UNCD/$ZrO_2$ and this phase was identified as graphite tribolayer [2]. From XRD analysis, densification and crystalline purity of grain is supposed to occur due to stress confinement. This claim is supported by the enhancement of diamond (111) and (220) plane intensity in the wear track. Other peaks are related to non-diamond phase and it is out of the scope of present work. Moreover, the shift in these XRD peaks to lower 2θ is indicative towards tribo-induced stress in wear track. Graphitic tribolayer locally formed possibly due to this tribo-induced stress during sliding and thus the stable low friction value is observed in UNCD/$Al_2O_3$ sliding combination.

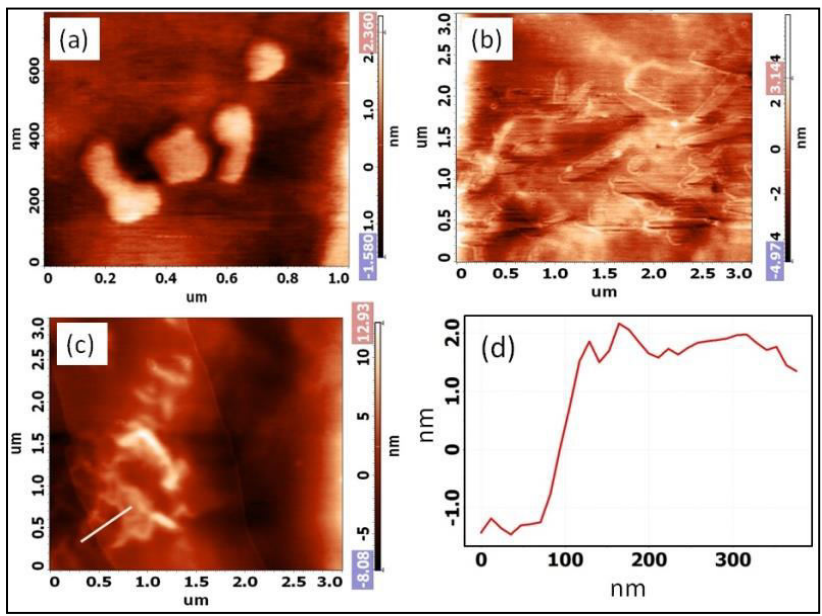

Fig. 4- (a) (b) and (c) AFM topography inside the wear track of UNCD/$Al_2O_3$ sliding case and (d) height profile at location mentioned by white line in panel (c); Tribology parameter of track: Load: 2 N, Sliding speed: 4 cm/sec, Sliding distance: 500m

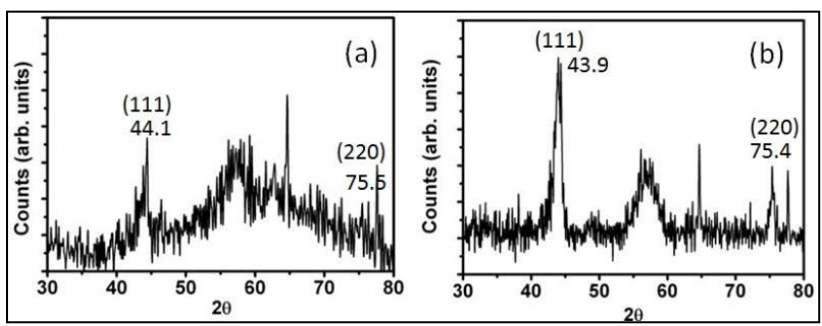

Fig. 5- (a) XRD analysis of UNCD film and (b) inside the wear track of UNCD/$Al_2O_3$ sliding combination; Tribology parameters of track: Load: 2 N, Sliding speed: 4 cm/sec, Sliding distance: 500m

## Conclusions

1. Tribological studies of this film were carried out against three different balls in ambient condition.
2. In the beginning of sliding process, friction coefficient for UNCD/steel sliding system was high but it was gradually decreased to low value with sliding process. This was attributed to oxidation mechanism.
3. High friction in UNCD/SiC system was governed by interfacial adhesion due to chemical bond formation.
4. Ultralow friction value was observed in the sliding combination of UNCD/$Al_2O_3$. In this case, such improvement was caused by chemically inert interfaces and formation of graphitic tribolayer which was determined by AFM and ECR measurements. The chemical structure of such layer was confirmed earlier [2].
5. The wear resistance of film was extremely high in all three cases with negligible wear loss.

## Acknowledgment

The authors would like to acknowledge Dr. K. Ganesan and Dr. S. K. Srivastava, IGCAR/Kalpakkam; Dr. D. Dinesh Kumar, Sathyabama University/Chennai; Mr. Pankaj Kr. Das, NIT/Agartala and Dr. S. Chakravarty, UGC-DAE CSR, Kalpakkam Node for various experimental helps.